\documentstyle[12pt]{article}

\def\pconst{2^{\fr{1}{4}}}
\def\mconst{ \Bigl( \frac{1}{2} \Bigr)^{\frac{1}{4}}}
\def\pd{\partial}
\def\bpd{{\bar \partial}}
\def\tpd{\partial_{\tau}}
\def\spd{\partial_{\sigma}}
\def\da{\dot{a}}
\def\db{\dot{b}}
\def\a{\alpha}
\def\b{\beta}
\def\balpha{{\bar \alpha}}
\def\bbeta{{\bar \beta}}
\def\s{\sigma}
\def\t{\tau}
\def\eps{\epsilon}
\def\veps{\varepsilon}
\def\th{\theta}
\def\lam{\lambda}
\def\bth{{\bar \theta}}

\def\beps{{\bar \epsilon}}
\def\Gm{\Gamma}
\def\gm{\gamma}
\def\momp{p^+}
\def\hdelta{{\hat \delta}}
\def\hDelta{{\hat \Delta}}
\def\bari{{\bar i}} 

\def\sq{\sqrt}
\def\twosq{\hbox{$\sqrt 2$}}
\def\e{{\rm e}}
\def\half{\frac{1}{2}}
\def\fr{\frac}

\def\pp{\prime}
\def\sint{\int^{\pi}_0 \fr{d\s}{i\pi}}
\def\bb{\begin{equation}}
\def\ee{\end{equation}}
\def\bba{\begin{eqnarray}}
\def\eea{\end{eqnarray}}

\begin{document}

\begin{titlepage}

\begin{tabbing}
   qqqqqqqqqqqqqqqqqqqqqqqqqqqqqqqqqqqqqqqqqqqqqq 
   \= qqqqqqqqqqqqq  \kill 
         \>  {\sc KEK-TH-504}    \\
         \>       hep-th/9612234 \\
         \>  {\sc December 1996} 
\end{tabbing}
\vspace{5mm}

\begin{center}
{\Large {\bf Vertex Operators for Super Yang-Mills \break 
 and Multi D-branes \break in Green-Schwarz Superstring}}
\end{center}

\vspace{1.5cm}

\centering{\sc Ken-ji Hamada}\footnote{E-mail address : 
hamada@theory.kek.jp}

\vspace{1cm}

\begin{center}
{\it National Laboratory for High Energy Physics (KEK),} \\
{\it Tsukuba, Ibaraki 305, Japan}
\end{center} 

\vspace{1cm}

\begin{abstract} 
We study vertex operators for super Yang-Mills and multi D-branes 
in covariant form using Green-Schwarz formalism. 
We introduce the contact terms naturally  and 
prove space-time supersymmetry and gauge invariance. 
The nonlinear realization of broken supersymmetry in the 
presence of D-branes is also discussed. 
The shift of fermionic coordinate   
$\delta^{(-)}\Psi =\eta$ becomes exact symmetry of D-brane 
in the static gauge, where $\eta$ is a constant spinor in $U(1)$ 
direction.       
\end{abstract}
\end{titlepage}

\section{Introduction}
\indent

   The D-brane description~\cite{dlp,p1,p2} of 
string solitons~\cite{dkl} has provided us   
with a powerful tool for studying world-volume theories of 
these extended objects in a perturbative formulation.   
Dp-branes are described in terms of open superstring theories  
that have Neumann boundary conditions in the $(p+1)$ directions 
and Dirichlet boundary conditions in the remaining $9-p$ 
directions which represent the locations of D-branes. 
The boundary relates left-moving and right-moving degrees of 
freedom so that it breakes half of space-time supersymmetry. 
Further, the Dirichlet boundary condition breaks the translational 
invariance of space-time. The Nambu-Goldstone modes associated 
with the broken symmetries are identified with  
collective coordinates of D-brane in the static gauge~\cite{hp}.       

  In this paper we construct  vertex operators of multi D-branes  
in covariant form using Green-Schwarz (GS) 
superstring~\cite{gs1,gs2,gsw} and 
prove that the supersymmetry of multi D-branes is given by that 
of reduced ten dimensional 
super Yang-Mills theory~\cite{w}.  
Multi D-branes are described by attaching the Chan-Paton factors 
on vertex operators.     
To construct non-abelian vertex operators in covariant form,  
we must introduce the contact terms to ensure   
supersymmetry and gauge invariance. The importance of the 
contact terms is stressed in the work by Green and Seiberg~\cite{gs}. 
Here we introduce the contact terms naturally and prove  
supersymmetry in the coordinate space.   
   
  This paper is organized as follows. In the next section 
we give a brief review of GS superstring. We then fix our 
notations and conventions. We first discuss the 
abelian case in Sect.3. We here give the vertex operators of  
D-brane in the coordinate space. We then generalize the 
argument to the non-abelian case in Sect.4 by replacing field 
strength and derivative with the covariant ones. 
This means that we introduce the contact terms like 
$[A_{\mu},A_{\nu}]$ naturally. 
Naively, the addition of the contact terms would break the 
supersymmetry-transformation law of vertex operators at the 
order of coupling constant $g$. We here show, however, that 
the contact terms coming from the collision points of vertex 
operators exactly cancel the extra terms that break 
the transformation law  and restore the supersymmetry.
In Sect.5 we discuss the nonlinear realization~\cite{gg95} of 
the broken supersymmetry in the presence of D-branes. 
We here develop the argument of 
Green and Gutperle formulated in 
closed-string/cylinder frame~\cite{gg}. 
The supersymmetry of D-branes becomes that of reduced  
super Yang-Mills theory, which is linearly realized by the 
unbroken supersymmetry $Q^+$ which satisfies the boundary state 
condition $Q^+ |B>=0$.  The nonlinear realization  
of broken supersymmetry satisfying $Q^- |B> \neq 0$ is given 
by a shift of fermionic collective coordinate,  
$\delta^{(-)}\Psi =\eta$, 
where $\eta$ is a constant spinor in $U(1)$ direction.

\section{The Set Up}
\setcounter{equation}{0}
\indent

 In this section we review the formulation of GS superstring 
and the procedure of light-cone gauge fixing. We then fix 
our notations and conventions. The covariant action of 
GS-superstring~\cite{gs1,bstt} is 
\bba
 S &=& -\fr{1}{\pi} \int d^2 \xi \biggl[ 
      \hbox{$\sq{ -h}$} 
       + i \eps^{ab} \pd_a X^{\mu}
             \Bigl( \bth^{(1)} \Gm_{\mu}\pd_b \th^{(1)} 
                     - \bth^{(2)} \Gm_{\mu}\pd_b \th^{(2)} \Bigr) 
           \nonumber \\  
   && \qquad\qquad\qquad\qquad
       -\eps^{ab} \bth^{(1)} \Gm^{\mu} \pd_a \th^{(1)} 
                  \bth^{(2)} \Gm_{\mu} \pd_a \th^{(2)} 
                       \biggr] ~ ,
\eea
where $\xi^a  =(\tau, \s)$, $\{ \Gm^{\mu},\Gm^{\nu} \} 
=-2\eta^{\mu\nu}$, 
$\eta^{\mu\nu}=(-1,1, \cdots, 1)$ and  $\bth = \th^t \Gm^0$.  
$h_{ab}=\Pi^{\mu}_a \Pi^{\nu}_b \eta_{\mu\nu}$ and 
$\Pi^{\mu}_a$ is defined by 
\bb
  \Pi^{\mu}_a =\pd_a X^{\mu}-i \bth^{(1)} \Gm^{\mu}\pd_a \th^{(1)} 
                     -i\bth^{(2)} \Gm^{\mu}\pd_a \th^{(2)} ~. 
\ee
The action is invariant under the global supersymmetry 
$\hdelta X^{\mu} =  \sum_A i \beps^{(A)} \Gm^{\mu} \th^{(A)} $,        
$\hdelta \th^{(A)} = \eps^{(A)} , ~(A=1,2) $ 
and also under the local fermionic symmetry ($\kappa$-symmetry)
$\hdelta X^{\mu} = \sum_A i \bth^{(A)} \Gm^{\mu} \a^{(A)}$, 
$\hdelta \th^{(A)} = \a^{(A)} $. 
$\a^{(A)}$'s are local fermionic fields defined by
$\a^{(A)} = \Bigl( 1-(-1)^A {\tilde \Gm} \Bigr) \kappa^{(A)} $,
where 
${\tilde \Gm}=\fr{1}{2\sq{-h}}\eps^{ab}{\tilde \Gm}_a 
{\tilde \Gm}_b $ and ${\tilde \Gm}_a = \Pi^{\mu}_a \Gm_{\mu}$.
The ${\tilde \Gm}$-matrix  satisfy the algebra 
$\{ {\tilde \Gm}_a, {\tilde \Gm}_b \}=-2h_{ab}$ and 
${\tilde \Gm}^2 =1$.

  The quantization of GS-superstring has been carried out in 
the light-cone gauge 
\bb
      X^+ = \momp \tau ~, \qquad \Gm^+ \th^{(A)}=0  ~,
\ee
where $X^{\pm}=\fr{1}{\sq 2}(X^0 \pm X^9)$ and 
$\Gm^{\pm}=\fr{1}{\sq 2}(\Gm^0 \pm \Gm^9)$. To preserve the 
gauge we have to take $\hdelta X^+ =0$ and 
$\Gm ^+ \hdelta \th^{(A)}=0$. This is carried out by combining 
the global supersymmetry and $\kappa$-symmetry with 
$\kappa^{(A)} =(-1)^{A-1} \fr{1}{2\momp}\pd_{\s}X^I 
\Gm^+ \Gm^I \eps^{(A)}$. We then get 
\bba
   \hdelta \th^{(1)} 
      &=& -\fr{1}{2\momp}\Gm^+ \Bigl\{ 
               (\pd_{\tau}- \pd_{\s})X^I \Gm^I -\momp\Gm^- 
                      \Bigr\} \eps^{(1)} ~, 
           \nonumber \\
   \hdelta \th^{(2)} 
      &=& -\fr{1}{2\momp}\Gm^+ \Bigl\{ 
               (\pd_{\tau}+ \pd_{\s})X^I \Gm^I -\momp\Gm^- 
                      \Bigr\} \eps^{(2)} ~, 
                 \\
   \hdelta X^{\mu} 
      &=& 2i \beps^{(1)} \Gm^{\mu} \th^{(1)} 
           + 2i \beps^{(2)} \Gm^{\mu} \th^{(2)} 
          + i \bth^{(1)} \Gm^{\mu} \hdelta\th^{(1)} 
          + i \bth^{(2)} \Gm^{\mu} \hdelta\th^{(2)} ~. 
            \nonumber 
\eea
These just satisfy the conditions $\hdelta X^+ = \Gm^+ \delta\th^{(A)}=0$ 
because of $(\Gm^+)^2=0$. 
Since $\bth^{(A)} \Gm^I \pd_a \th^{(A)} =0$ in the light-cone 
gauge, the action reduces to the simple form
\bba
  S_{LC} &=&
         -\fr{1}{2\pi}\int d^2 \xi \Bigl[ 
        -\bigl( \tpd X^I \bigr)^2 
          +\bigl( \spd X^I \bigr)^2 
         -2i \momp \bth^{(1)} \Gm^- (\tpd +\spd)\th^{(1)} 
           \nonumber \\         
   && \qquad\qquad\qquad\qquad\qquad
          -2i \momp \bth^{(2)} \Gm^- (\tpd -\spd)\th^{(2)} 
              \Bigr] ~.
\eea 

  In the following we only consider chiral theory with negative 
chirality, $\Gm^{11}\th^{(A)} = -\th^{(A)}$. The extension to 
non-chiral theory is straightforward. Let us introduce the variables
\bb
    S^{(A)} =  \pconst i \hbox{$\sq{\momp}$}\Gm^- \th^{(A)} ~, 
      \qquad \hbox{or} \qquad 
       \th^{(A)} =\half \fr{1}{\pconst i\sq{\momp}}\Gm^+ S^{(A)} 
\ee 
such that $\Gm^{11}S^{(A)} =S^{(A)}$ and $\Gm^- S^{(A)} =0$. 
The action is then written in the form
\bba
  S_{LC} &=&
         -\fr{1}{2\pi}\int d^2 \xi \Bigl[ 
        -\bigl( \tpd X^I \bigr)^2 
          +\bigl( \spd X^I \bigr)^2 
         -i \bigl( S^{(1)} \bigr)^t (\tpd +\spd)S^{(1)} 
           \nonumber \\         
   && \qquad\qquad\qquad\qquad\qquad
          -i\bigl(  S^{(2)} \bigr)^t  (\tpd -\spd) S^{(2)} 
              \Bigr] ~.
\eea 
In this paper we use the following $\Gm$-matrix;  
$
   \Gm^0 = i\left( \begin{array}{cc}
                    0 & -I  \\
                    I & 0 
                  \end{array} \right) 
$, 
$
   \Gm^9 = i\left( \begin{array}{cc}
                    0 & J  \\
                    J & 0 
                  \end{array} \right) 
$, 
$
   \Gm^I = i\left( \begin{array}{cc}
                    0 & {\hat \gm}^I  \\
                    {\hat \gm}^I & 0 
                  \end{array} \right) 
$
and
$
   \Gm^{11} = \left( \begin{array}{cc}
                    I & 0  \\
                    0      & -I 
                  \end{array} \right) 
$,  
where $I$ is $16 \times 16$ identity matrix,  
$
   J = \left( \begin{array}{cc}
                    I_8 & 0  \\
                    0   & -I_8
                  \end{array} \right) 
$ 
and 
$
   {\hat \gm}^I = \left( \begin{array}{cc}
                    0 & \gm^I  \\
                    (\gm^I)^t & 0 
                  \end{array} \right) 
$. $I_8$ is $8 \times 8$ identity and $\gm^I$ is $8 \times 8$ 
$\gm$-matrix 
familiar in the light-cone gauge formulation of 
GS-superstring~\cite{gsw,gs1,gs2}. 
Hence $(S^{(A)})^t = (S^{(A)a},0,0,0), ~a=1,\cdots, 8$.

  In $SO(8)$ $\gm$-matrix notation the supersymmetry 
transformation becomes
\bba
  &&   \hdelta^{(A)}_{\eps} X^I 
            = 2i{\bar \eps}^{(A)}\Gm^I\th^{(A)} 
            = -\pconst \fr{i}{\sq{\momp}}
                 \eps^{(A)\da}\gm^I_{\da a}S^{(A)a} ~, 
               \\
  &&   \hdelta^{(A)}_{\eps} S^{(A)} = 
              \pconst \hbox{$\sq{2\momp}$} \eps^{(A)a} 
              +\pconst \fr{1}{\sq{\momp}} 
               \eps^{(A)\da}\gm^I_{\da a}\tpd X^I ~,  
\eea
where $\Gm^{11}\eps^{(A)}=-\eps^{(A)}$ and  
$\hdelta^{(A)}_{\eps}f =[ \eps^{(A)a} Q^{(A)a} + 
\eps^{(A)\da}Q^{(A)\da}, f ]$. 
The transformation law of $\hdelta_{\eps}X^- $ can read out 
from eq.(2.4). 
The supercharges are defined by 
\bba
  && Q^{(1)a} =\pconst \hbox{$\sq{2\momp}$}
              \int^{\pi}_0 \fr{d\s}{\pi} S^{(1)a}(\s)~, 
         \qquad 
   Q^{(2)a} =\pconst \hbox{$\sq{2\momp}$}
               \int^{\pi}_0 \fr{d\s}{\pi} S^{(2)a}(\s)~, 
             \nonumber \\
  && \qquad\qquad
         Q^{(1)\da} =\pconst \fr{2}{\sq{\momp}}
               \int^{\pi}_0 \fr{d\s}{\pi} 
                 \pd X^I(\s) S^{(1)a}(\s) \gm^I_{a\da} ~, 
         \\
  && \qquad\qquad
       Q^{(2)\da} = \pconst \fr{2}{\sq{\momp}}
               \int^{\pi}_0 \fr{d\s}{\pi} 
                 \bpd X^I (\s) S^{(2)a}(\s)\gm^I_{a\da} ~. 
                \nonumber 
\eea
where $\pd= \half (\tpd -\spd)$ and $\bpd=\half (\tpd +\spd)$.
These satisfy the superalgebra 
$\{ Q^{(A)},Q^{(B)t} \}=2p^{\mu}\Gm_{\mu}C  \delta^{AB}$, where 
$Q^{(A)t} =(Q^{(A)a},Q^{(A)\da},0,0)$ and $C=-\Gm^0$ is the 
charge conjugation matrix.

\section{Vertex Operators for D-branes}
\setcounter{equation}{0}
\indent

 We first discuss the vertex operators for a single Dp-brane. 
The Dp-brane configuration is described by the boundary 
conditions 
\bba
   && \Bigl( \pd X^I(\t,\s) -N_{IJ}{\bar \pd}X^J(\t,\s)  
                       \Bigr)\vert_{\s=0,\pi}=0 ~,
               \\
   && \Bigl( S^{(1)a}(\t,\s) -N_{ab}S^{(2)b}(\t,\s) 
                       \Bigr)\vert_{\s=0,\pi}=0 ~. 
\eea
Here we take $N_{IJ}=(\delta_{\a\b}, -\delta_{ij})$ so that   
$\a,\b=1,\cdots, p-1$ are the Neumann directions and $i,j=p,\cdots, 8$ 
are the Dirichlet directions.  
$N_{ab}$ satisfies the orthogonal 
condition $N_{ac}N_{bc}=\delta_{ab}$. 
$p$ is odd because we now consider the chiral theory. 
When we consider the non-chiral theory, $p$ must be even.  
$p=9$ is the usual open superstring. Since $X^{\pm}$'s satisfy 
Neumann boundary conditions, the value of $p$ is restricted 
to $p \geq 1$. To discuss the case of $p <1$, we must  
go to the cylinder frame which is discussed in the Sect.5.

 The boundary conditions break half of the supersymmetry. 
The left-moving and the right-moving modes are related by the 
boundary conditions so that  
$\pd X^I(\t,\s) -N_{IJ}{\bar \pd}X^J(\t,-\s)=0 $ and 
$ S^{(1)a}(\t,\s) -N_{ab}S^{(2)b}(\t,-\s) =0 $ are satisfied. 
Therefore only the supercharges 
\bb
   Q^a =\fr{1}{\sq{2}}
          \bigl( Q^{(1)a} +N_{ab}Q^{(2)b} \bigr) ~, 
             \qquad  
   Q^{\da} =\fr{1}{\sq{2}} 
            \bigl( Q^{(1)\da} +N_{\da\db}Q^{(2)\db} \bigr)~,
\ee 
survive, where $N_{\da\db}$ satisfies the orthogonal condition 
and 
\bb
      \gm^I_{a\da}N_{IJ}- N_{ab}N_{\da\db}\gm^J_{b\db}=0 ~.
\ee
The solution is given by $N_{ab}= (\gm^p \cdots \gm^8)_{ab}$ 
and $N_{\da\db}= (\gm^p \cdots \gm^8)_{\da\db}$.   
The supercharges satisfy the N=1 superalgebra $\{ Q,Q^t \}
=2p^{\mu}\Gm_{\mu}C $ or
\bb
  \{ Q^{a},Q^{b} \}=2\twosq\momp \delta^{ab} ~, \quad 
  \{ Q^{a},Q^{\db} \}=2 \gm^{\a}_{a\db}~p^{\a} ~, \quad 
  \{ Q^{\da},Q^{\db} \}=2\twosq p^- \delta^{\da\db}~,
\ee
where
\bb
   p^-= \fr{1}{2\momp} \int^{\pi}_0 \fr{d\s}{\pi}  
          \Bigl[ \bigl( {\dot X}^I \bigr)^2 
                  +\bigl( X^{\pp I} \bigr)^2
           -i S^{(1)a} \spd S^{(1)a} 
              +i S^{(2)a} \spd S^{(2)a} \Bigr] ~.
\ee
The supersymmetry transformation at the boundary is now given  
in the form
\bba
  && \hdelta_{\eps} X^{\a}
      = -\pconst \fr{i}{\sq{\momp}}
        \eps^{\da}\gm^{\a}_{\da a}S^{a} ~, 
   \qquad 
     \hdelta_{\eps} X^i = 0 ~, 
          \nonumber \\
  && \quad
        \hdelta_{\eps} S^{a} 
       = \pconst \hbox{$\sq{2\momp}$}\eps^a 
          +\pconst \fr{1}{\sq{\momp}}\eps^{\da}
                    \gm^I_{\da a}\pd_B X^I ~,
\eea
where $S^a =\fr{1}{\sq{2}}(S^{(1)a}+N_{ab}S^{(2)b})$ 
and $\pd_B X^I =\pd X^I +N_{IJ}\bpd X^J 
=(\tpd X^{\a}, \break -\spd X^i)$.  

 Let us intoduce the 10D superfield ${\cal A}_{\mu}(X,\th)$ 
defined by
\bba
     {\cal A}_{\mu}(X,\th)&=& \e^{\bth G}A_{\mu}(X)\e^{-\bth G}
                \nonumber \\
                   &=& A_{\mu}(X) +\delta_{\th}A_{\mu}(X) 
                      +\fr{1}{2!}\delta^2_{\th}A_{\mu}(X)+ \cdots ~,
\eea
where $\th= (\pconst 2i \sq{\momp})^{-1}\Gm^+ S$. 
$G$ is a generator of space-time supersymmetry.
The vertex operator $V({\cal A}_{\mu}(X,\th))$ should satisfy the 
following relation;
\bb
   \hdelta_{\eps}V({\cal A}_{\mu}(X,\th)) 
      = V(\delta_{\eps}{\cal A}_{\mu}(X,\th)) ~.
\ee
We will show that this equation is satisfied  when the vertex 
operator is given by 
\bb 
      V({\cal A}(X,\th))
          = \int d\tau {\cal A}_{\mu}(X,\th)
                             \pd_B X^{\mu}~, 
\ee
where $\pd_B X^{\mu}=(\tpd X^+, \tpd X^- , \pd_B X^I )$ 
and the space-time supersymmetry is given by 
\bb   
    \delta_{\eps}A^{\mu}(X) 
           = -i{\bar \Psi(X)}\Gm^{\mu}\eps ~,
       \qquad
    \delta_{\eps}\Psi(X)= -\half F_{\mu\nu}(X)\Gm^{\mu\nu}\eps 
\ee
 and 
\bba
  &&  A^{\mu}= \Bigl( A^{\balpha}(X^{\balpha})~, 
                  ~\phi^i (X^{\balpha}) \Bigr) ~,
               \\
  &&  F_{\mu\nu}=\pd_{\mu}A_{\nu} -\pd_{\nu}A_{\mu}~,  
\eea
where ${\bar \a} =(+,-,\a)$. $\phi^i$ denotes the collective 
coordinate of D-brane in the ``static gauge''.  
The fields depend only on the coordinates 
$X^{{\bar \a}}$ such that $\pd^i A^{\mu}=\pd^i \Psi=0$  
and then $ F_{{\bar \a} i}=-F_{i{\bar \a}}=\pd_{\bar \a}\phi^i$ 
and $F_{ij}=0 $. These fields satisfy the equations of motion
$\pd^{\bar \a}F_{{\bar \a}\mu}= 0$ and 
$\Gm^{\bar \a}\pd_{\bar \a}\Psi=0$. $\Psi$ has the negative 
chirality, $\Gm^{11}\Psi=-\Psi$. 
The explicit form of the vertex operator is now given in the form
\bba
   V({\cal A})&=& V_B(A) + V_F(\Psi) ~, 
                     \\ 
     V_B(A)&=& \int d\tau \biggl[ 
              A_{\mu}(X) 
              -\fr{i}{2\cdot 2!}F_{\nu\lam}(X)
                (\bth \Gm_{\mu}\Gm^{\nu\lam}\th)
            \nonumber \\ 
          && \quad\qquad 
              -\fr{1}{2\cdot 4!}\pd_{\nu}F_{\lam\rho}(X) 
                (\bth\Gm_{\kappa}\Gm^{\lam\rho}\th) 
                (\bth\Gm_{\mu}\Gm^{\nu\kappa}\th) 
                 \biggr] \pd_B X^{\mu} ~,
                     \\ 
   V_F(\Psi) &=& \int d\tau \biggl[ 
               -i{\bar \Psi}(X)\Gm_{\mu}\th  
               -\fr{1}{3!}\pd_{\nu}{\bar \Psi}(X)\Gm_{\lam}\th  
                (\bth\Gm_{\mu}\Gm^{\nu\lam}\th)  
               \nonumber \\
            && \qquad
               +\fr{i}{5!}\pd_{\nu}\pd_{\lam}{\bar \Psi}(X) 
                \Gm_{\rho}\th 
                    (\bth\Gm_{\kappa}\Gm^{\lam\rho}\th) 
                    (\bth\Gm_{\mu}\Gm^{\nu\kappa}\th) 
                       \biggr] \pd_B X^{\mu}~.
\eea
In the familiar $SO(8)$ $\gm$-matrix notation, these are rewritten  
in the form
\bba
   V_B(A) &=& \int d\tau \biggl[ 
                 A^I \pd_B X^I -A^+ \tpd X^- -A^- \momp
              -\fr{i}{8}F_{IJ}(S\gm^{IJ}S) 
               \nonumber \\
          && \quad 
              -\fr{i}{4\momp}F^{+I}(S\gm^{IJ}S) \pd_B X^J 
              -\fr{1}{96\momp}\pd^J F^{+I}(S\gm^{IK}S)
                    (S\gm^{JK}S)
                \nonumber \\
          && \qquad\qquad\quad
              +\fr{1}{96 p^{+2}}\pd^+F^{+I}(S\gm^{IK}S)
                    (S\gm^{JK}S) \pd_B X^J  \biggr] ~, 
                      \\ 
    V_F(\Psi) &=& i \mconst \int d\tau \biggl[ 
            \hbox{$\sq{\momp}$}(\psi S)  
             +\fr{1}{\sq{2\momp}}(\psi\gm^I S)\pd_B X^I 
                \nonumber \\ 
          && \qquad\qquad
               -\fr{i}{12\sq{2\momp}} (\pd^I \psi\gm^J S)    
                      (S\gm^{IJ}S) 
               \nonumber \\
          && \quad
               +\fr{i}{12\momp\sq{2\momp}}
                 (\pd^+ \psi \gm^J S)(S\gm^{IJ}S) \pd_B X^I
                          \\
          && \quad
               - \fr{1}{240 \momp\sq{2\momp}} 
                   ( \pd^J \pd^+ \psi \gm^I S) 
                   (S\gm^{IK}S)(S\gm^{JK}S) 
                 \nonumber \\ 
          && \quad
               + \fr{1}{240 p^{+2}\sq{2\momp}} 
                   ( \pd^+ \pd^+ \psi \gm^I S) 
                   (S\gm^{IK}S)(S\gm^{JK}S) \pd_B X^J  
                        \biggr]~, 
                  \nonumber 
\eea
where we use the notations $\Psi=(0,0,\psi^a,\psi^{\da})$ and 
$S\gm^{IJ}S =S^a \gm^{IJ}_{ab} S^b$, 
$\psi S=\psi^a S^a$ and $\psi \gm^I S= \psi^{\da}\gm^I_{\da a}S^a$.

  The proof of supersymmetry in general case is rather 
complicated. In this section we work in the case 
$\pd^+ A^{\mu}=\pd^+ \Psi =0$, but $A^{\pm} \neq 0$.     
Using the supersymmetry transformation (3.7) and noting 
$\hdelta_{\eps}f(X)=\pd_{\mu}f(X)\hdelta_{\eps} X^{\mu}~
=\pd^{\a}f(X)\hdelta X^{\a}$, where   
$\pd^+ f=\pd^i f= \hdelta_{\eps} X^+ =0$ are used, 
and also 
\bb     
   \hdelta_{\eps}(\pd_B X^I) 
           = \tpd \bigl( \hDelta_{\eps} X^I \bigr)
           = -\pconst \fr{i}{\sq{\momp}}
              \eps^{\da}\gm^I_{\da a} \tpd S^a ~,
\ee 
we obtain the following equations;
\bba
     \hdelta_{\eps}{\cal V}_B(A) &=& {\cal V}_F(\delta_{\eps}\Psi) 
              + \tpd \Bigl( 
                     A_{\mu}{\hat \Delta}_{\eps}X^{\mu} 
                              \Bigr)~,
                   \\ 
     \hdelta_{\eps}{\cal V}_F(\Psi) &=& {\cal V}_B(\delta_{\eps}A) 
              + \tpd \biggl(  
                   -\fr{1}{8\sq{2}}\fr{1}{\momp}
                     (\psi^{\da} \gm^{IJ}_{\da\db}\eps^{\db})
                      (S\gm^{IJ}S) \biggr)~, 
\eea
where 
\bb 
         V_{B,F}=\int d\tau {\cal V}_{B,F}
\ee
and $A_{\mu}\hDelta_{\eps}X^{\mu}=A^{\a} 
\hdelta_{\eps}X^{\a}-A^+ \hdelta_{\eps} X^- 
+\phi^i \hDelta_{\eps} X^i$.  
Note that $\hDelta_{\eps}X^{\a}=\hdelta_{\eps}X^{\a}$ but 
$\hDelta_{\eps} X^i \neq \hdelta_{\eps} X^i ~(=0)$.  
The $\tau$-derivative terms play an 
important role when we discuss the non-abelian 
gauge field~\cite{gs}.

\section{Vertex Operators for Multi D-branes}
\setcounter{equation}{0}
\indent

   In this section we study the extension of the previous 
arrgument to the non-abelian case. It is now straightforward. 
We replace the gauge field and the fermionic field with  
$A_{\mu}=A^s_{\mu}\lam^s$ and $\Psi=\Psi^s \lam^s$, 
where $\lam^s$ is 
the generator of gauge group. We also replace 
the field strength (3.13) with  
\bb
     F_{\mu\nu}= \pd_{\mu}A_{\nu}-\pd_{\nu}A_{\mu}
                    +ig^{\pp}~[A_{\mu},A_{\nu}] 
\ee
and the space-time derivative $\pd_{\mu}$ with the covariant 
one  
\bb
      D_{\mu}= \pd_{\mu} +ig^{\pp}~[ A_{\mu}, \quad] ~. 
\ee
The fields satisfy the equations of motion
$ D^{\mu}F_{\mu\nu} =0 $ and $ \Gm^{\mu}D_{\mu}\Psi =0 $. More 
explicitly, 
\bba
   &&   D^{\balpha}F_{\balpha\bbeta} 
         +ig^{\pp}[\phi^i , F_{i\bbeta}]=0 ~, \quad 
        D^{\balpha}F_{\balpha j} 
         +ig^{\pp}[\phi^i , [\phi_i ,\phi_j] ]=0
              \\
  && \qquad\qquad\qquad  \Gm^{\balpha}D_{\balpha}\Psi 
           +ig^{\pp}\Gm^i [\phi_i , \Psi] =0 ~.
\eea
where $F_{\balpha i}=\pd_{\balpha}\phi_i 
+ig^{\pp}[A_{\balpha}~,~\phi_i]$.
This means that the Chan-Paton factor is attached at the boundary. 
The thing we must take care here is the treatment of the contact 
term $[A_{\mu},A_{\nu}]$. We here define 
the contact term, in the momentum space, by 
\bb  
    [A_{\mu}(X),A_{\nu}(X)]=[\lam^s,\lam^t] \int dk_1 \int dk_2   
         A^s_{\mu}(k_1)A^t_{\nu}(k_2)\e^{i(k_1+k_2)\cdot X} ~, 
\ee
where $k \cdot X= k_{\balpha}X^{\balpha}$ and the 
integrals are restricted to $k_1\cdot k_2 =0$. 
In the below we will show that the following supersymmetry 
transformation is realized when $g^{\pp}=g$;
\bba
   && {\cal P} \int d\tau \Bigl( \hdelta_{\eps}{\cal V}_B(A)
                     +\hdelta_{\eps}{\cal V}_F(\Psi) \Bigr) 
         \exp \biggl\{ -ig\int d\tau 
             \Bigl( {\cal V}_B(A)+{\cal V}_F(\Psi) 
                     \Bigr) \biggr\} 
                \\ 
   && 
      = {\cal P} \int d\tau \Bigl( {\cal V}_B(\delta_{\eps}A)
                     +{\cal V}_F(\delta_{\eps}\Psi) \Bigr) 
         \exp \biggl\{ -ig\int d\tau 
             \Bigl( {\cal V}_B(A)+{\cal V}_F(\Psi) 
                               \Bigr) \biggr\} ~,
                   \nonumber 
\eea
where $\delta_{\eps}$ is the non-abelian supersymmetry defined in 
the form (3.11) with the field strength (4.1). 
${\cal P}$ is the $\tau$-ordering of  vertex operators.

  We first discuss the transformation property of $V_B(A)$. 
For simplicity we will show the supersymmetry in the background 
$A^+ =0$ and $\pd^+=0$. In this case the vertex operator  
reduces to the simple form
\bb
     V_B(A)= \int d\tau \biggl[ 
               A^I \pd_B X^I -A^- \momp 
                -\fr{i}{8}F^{IJ}(S\gm^{IJ}S) \biggr] ~.         
\ee 
As in the abelian case, noting that eq.(3.19) and  
$\hdelta_{\eps} A^I = \pd^{\a}A^I \hdelta_{\eps} X^{\a}= 
\pd^J A^I \hDelta_{\eps} X^J$ etc, the variation 
 $\hdelta_{\eps} {\cal V}_B(A)$ is calculated as 
\bba
      \hdelta_{\eps}{\cal V}_B(A) &=& 
      \Bigl( \pd^I A^J - \pd^J A^I \Bigr) 
                \pd_B X^J \hDelta_{\eps} X^I  
         - \Bigl( \pd^I A^- -\pd^- A^I \Bigr)\momp \hDelta_{\eps}X^I  
              \nonumber \\
  && -\fr{i}{8} \pd^K F^{IJ} \hDelta_{\eps} X^K 
       -\fr{i}{4}F^{IJ}(\hdelta_{\eps}S \gm^{IJ}S) 
       + \tpd \Bigl( A^I \hDelta_{\eps}X^I \Bigl)
              \nonumber \\
  &=&  F^{IJ}\pd_B X^J \hDelta_{\eps}X^I
        -F^{I-}\momp \hDelta_{\eps} X^I 
        -\fr{i}{8}D^{K}F^{IJ}(S\gm^{IJ}S)\hDelta_{\eps}X^K 
            \nonumber \\
   && -\fr{i}{4}F^{IJ}(\hdelta_{\eps}S \gm^{IJ}S) 
       -ig^{\pp}Y 
       + \tpd \Bigl( A^I \hDelta_{\eps}X^I \Bigl) ~,
\eea
where
\bb
    Y=  \Bigl( [A^I,A^J] \pd_B X^J  
          -[A^I,A^-] \momp \Bigr) \hDelta_{\eps} X^I
          -\fr{i}{8} [A^K,F^{IJ}](S\gm^{IJ}S)
                   \hDelta_{\eps} X^K  ~.
\ee
Applying the Fierz transformation and using the equation of motion 
$D^I F^{IJ}=0$ and Bianchi identity $D^{[K}F^{IJ]}=0$, 
we obtain the following equation;
\bb
    \hdelta_{\eps}{\cal V}_B(A)
       = {\cal V}_F(\delta_{\eps}\Psi)  -ig^{\pp} Y 
              +\tpd \Bigl( A^I \hDelta_{\eps}X^I \Bigr)  ~,
\ee
Using this we evaluate  
\bba
  && {\cal P} \int d\tau \hdelta_{\eps}{\cal V}_B(A) 
         \exp \biggl( -ig \int d\tau {\cal V}_B(A) \biggr)  
                \\
  && = \int d\tau \hdelta_{\eps}{\cal V}_B(A) 
       -ig {\cal P} \int d\tau^{\pp} \int d\tau 
             \hdelta_{\eps}{\cal V}_B(A(\tau^{\pp})) 
              {\cal V}_B(A(\tau)) + o(g^2) ~.
                \nonumber 
\eea
We will show that the extra term $Y$ which appears in the first 
term of r.h.s. cancels the contact term coming from the 
collision point of operators in the second term 
when $g^{\pp}=g$.  Noting the $\tau$-ordering of 
vertex operators, the contact term is evaluated in the following. 
The quantity which contributes to the contact term is 
\bba
  && -ig \biggl( \int d\tau 
                 \int^{\infty}_{\tau +\veps}d\tau^{\pp}
         \pd_{\tau^{\pp}} (A^I(\tau^{\pp}) \hDelta_{\eps} X^I ) 
             {\cal V}_B(A(\tau)) 
             \nonumber \\ 
  && \qquad\qquad\qquad
      + \int d\tau \int^{\tau -\veps}_{-\infty}d\tau^{\pp}
             {\cal V}_B(A(\tau))     
         \pd_{\tau^{\pp}} (A^I(\tau^{\pp}) \hDelta_{\eps} X^I ) 
                 \biggr) 
             \nonumber \\
  &&  =ig \int d\tau \int dk^{\pp} dk ~\veps^{k^{\pp} \cdot ~k} 
          \Bigl[ A^I(k^{\pp})~,~
             A^J(k) \pd_B X^J -A^-(k)\momp 
              \nonumber \\
    && \qquad\qquad\qquad
              -\fr{i}{8}F^{JK}(k)(S\gm^{JK}S) \Bigr]     
         \e^{i(k^{\pp} +k)\cdot X(\tau)} 
             \hDelta_{\eps}X^I(\tau) ~, 
\eea
where $\veps$ is a short distance cut-off. 
We here evaluate the singular part by going to the momentum 
space $f(X(\tau)) =\int dk f(k) \e^{ik\cdot X(\tau)}$ as follows;
\bb
    f_1(X(\tau^{\pp}))f_2(X(\tau))
     =\int dk^{\pp} dk f_1(k^{\pp})f_2(k)
         \e^{i(k^{\pp}+k)\cdot X(\tau)} 
          \vert \tau^{\pp} -\tau \vert^{k^{\pp} \cdot k} ~.
\ee 
When $k^{\pp} \cdot k =0$, the expression (4.12) gives a finite 
contribution at the limit $\veps \rightarrow 0$. 
This is just the contact term we want, which becomes 
$ig Y$ in the coordinate space. 
Thus, when $g^{\pp} =g$, the extra term $-ig^{\pp}Y$ in (4.10) 
exactly cancels the contact term derived from the second 
term of r.h.s. of (4.11). 
When $k^{\pp} \cdot k > 0$, (4.12) vanishes.  
It, however, becomes infinite when $k^{\pp} \cdot k <0$. 
In this case another kind of contact term is needed to subtract 
the infinity, which is not discussed in this paper. 
Resumming up in $g$ we thus obtain 
\bb
     {\cal P} \hdelta_{\eps}V_B(A) \exp \Bigl( -igV_B(A) \Bigr) 
  = {\cal P} V_F(\delta_{\eps}\Psi)\exp \Bigl( -igV_B(A) \Bigr) ~. 
\ee
In the following we set $g^{\pp}=g$. 

  Next we consider the transformation property of $V_F(\Psi)$. 
To preserve the gauge $A^+=0$ we introduce the gauge parameter 
$\Lambda = i{\bar \chi} \Gm^+ \eps 
=i\twosq \chi^{\da}\eps^{\da}$ as
\bb
    \delta^{\chi}_{\eps} A^{\mu} 
        = -i{\bar \Psi} \Gm^{\mu}\eps + D^{\mu} \Lambda ~. 
\ee
{}From $\delta A^+=0$, $\chi$ is given by 
${\bar \Psi} \Gm^+ \eps =\pd^+ {\bar \chi} \Gm^+ \eps$. 
Since we now consider the case $\pd^+ =0$, this means that 
we must take the background $\Gm^+ \Psi =\psi^{\da}=0$. 
Therefore, in this case, the vertex operator $V_F(\Psi)$  
reduces to the simple form
\bb
     V_F(\Psi)= i \mconst \int d\tau 
                 \hbox{$\sq{\momp}$}(\psi^a S^a) ~. 
\ee
In the following we carry out the calculation keeping the 
gauge parameter $\chi$ arbitrary. This means that we  
consider the background $A^+ =\psi^{\da}=0$ and also
$\delta A^+ =0$ because of $\pd^+ =0$, but 
$\delta \psi^{\da}\neq 0$ and $\delta^2 A^+ \neq 0$. 
When we want to preserve the $A^+=0$ gauge exactly we must take 
$ \chi^{\da}= \fr{1}{\pd^+}\psi^{\da} ~(\neq 0) $. 

   The supersymmetry transformation is now 
\bba
   &&  \hdelta_{\eps} {\cal V}_F(\Psi) 
       = i(\psi \gm^I \eps) \pd_B X^I 
          +\fr{1}{4}(\pd^I \psi \gm^J \eps)(S\gm^{IJ}S) 
              \nonumber \\
   && \qquad\qquad\qquad
        + i\fr{g}{16}[ A^I, \psi \gm^I \gm^{JK}\eps] 
                  (S\gm^{JK}S) ~,
\eea
where we use the equation of motion $D^I \psi \gm^I =0$. 
Using the expression of $V_B(A)$ (4.7) and  
$\delta^{\chi}_{\eps}A^{\mu}$ (4.15), it is rewritten in the form 
\bb
   \hdelta_{\eps}{\cal V}_F(\Psi) =
         {\cal V}_B(\delta^{\chi}_{\eps} A) +ig Z 
          -\tpd (i\twosq \chi\eps) ~,
\ee
where $\chi\eps=\chi^{\da} \eps^{\da}$. 
The extra term $Z$ is given by 
\bba
  && Z= \fr{1}{16}[A^I, \psi\gm^{JK}\gm^I\eps]
                 (S\gm^{JK}S)
         -\fr{1}{4\sq{2}} [F^{IJ},\chi\eps]
                  (S\gm^{IJ}S)  
               \nonumber \\
  && \qquad\qquad\qquad
       -i\twosq  \Bigl( [A^I, \chi\eps]\pd_B X^I 
                -[A^-, \chi\eps]\momp \Bigr) ~.
\eea  
The $\tau$-derivative term needs to cancel that appears 
in the expression $V_B(\delta^{\chi}_{\eps}A)$ when we 
incorporate the gauge transformation into the supersymmetry 
transformation as in (4.15).      
Since $\psi^{\da}=0$, the $\tau$-derivative term appearing 
in the expression (3.21) vanishes now. 

  Let us calculate the following quantity;
\bb
    {\cal P} \Bigl(\hdelta_{\eps}V_B(A)
               +\hdelta_{\eps}V_F(\Psi) \Bigr) 
           \exp \Bigl\{ -ig \Bigl( V_B(A)+V_F(\Psi) 
                              \Bigr) \Bigr\} 
\ee
As in the previous calculation the contact terms coming from 
the expression 
\bb
         -ig{\cal P} \int d\tau \tpd 
                (A^I \hDelta_{\eps}X^I )V_F(\Psi)
         -ig{\cal P} \int d\tau \tpd (-i\twosq \chi\eps )V_B(A)  
\ee        
just cancel $ig Z$. The contact term coming from 
$-ig {\cal P} \hdelta_{\eps}V_F(\Psi) V_F(\Psi)$ gives the 
extra term 
\bb
     -ig \mconst \hbox{$\sq{2\momp}$}
        \int d\tau [ \chi\eps, \psi S] ~.
\ee
Note that 
\bb
    V_F(\delta^{\chi}_{\eps} \Psi) 
         = V_F(\delta_{\eps} \Psi) 
             -ig \mconst \hbox{$\sq{2\momp}$}
                 \int d\tau [ \chi\eps, \psi S] ~,
\ee
where $\delta^{\chi}_{\eps}$ is defined by 
$
     \delta^{\chi}_{\eps} \Psi 
       = \delta_{\eps} \Psi -ig[\Lambda, \Psi] 
$
with $\Lambda =i{\bar \chi}\Gm^+ \eps$. Thus the extra 
contact term (4.22) and $V_F(\delta_{\eps}\Psi)$ in (4.14) are 
combined into $V_F(\delta^{\chi}_{\eps} \Psi)$ and we finally 
obtain 
\bba
  &&  {\cal P} \Bigl(\hdelta_{\eps}V_B(A)
               +\hdelta_{\eps}V_F(\Psi) \Bigr) 
           \exp \Bigl\{  -ig \Bigl( V_B(A)+V_F(\Psi) 
                             \Bigr) \Bigr\} 
              \nonumber \\
  &&  =  {\cal P} \Bigl( V_B(\delta^{\chi}_{\eps}A) 
             + V_F (\delta^{\chi}_{\eps} \Psi) \Bigr) 
        \exp \Bigl\{ -ig \Bigl( V_B(A)+V_F(\Psi) 
                          \Bigr) \Bigr\} ~.
\eea

  Until now we assumed the gauge invariance in the calculation. 
Since the equation is satisfied for the generic $\chi$, however,  
this also gives the partial proof of the gauge invariance. 
The proof for the generic background is rather complicated, but 
straightforward.

\section{Nonlinear Realized Supersymmetry of D-branes}
\setcounter{equation}{0}
\indent

  In this section we study the D-brane in the 
cylinder/closed-string frame and then discuss the nonlinear 
realization of the broken supersymmetry. 
We introduce the boundary states 
satisfying the boundary condition~\cite{gg} 
\bba
   &&   \Bigl( \pd X^I (\s) -M_{IJ} \bpd X^J (\s)\Bigr)
                  \vert B >=0 ~,
               \\ 
   &&   \Bigl(  S^{(1)a}(\s) +i M_{ab}S^{(2)b}(\s)\Bigr) 
                   \vert B > =0 ~.
\eea
at the boundary $\tau=0$, 
where $I=(\a=1, \cdots, p+1 ~,~i=p+2, \cdots, 8)$ and 
$M_{IJ}= (-\delta_{\a\b} , \delta_{ij})$ and 
$M_{ac}M_{bc}=\delta_{ab}$. $\a,\b$ denote the Neumann boundary 
condition and $i,j$ is Dirichlet ones. The boundary condition of 
coordinates $X^{\pm}$ now become the Dirichlet ones so that the 
value of $p$ is restrited to $-1 \leq p \leq 7$ in the cylinder 
frame. This means that the 
world-volume has the Euclidean signature and one of the Dirichlet 
directions is time-like. So it is related to the D-brane by a 
double Wick rotation~\cite{gg}. 

  Let us define the supercharges 
\bb
   Q^{\pm a} =\fr{1}{\sq{2}}
        \bigl( Q^{(1)a} \pm iM_{ab}Q^{(2)b} \bigr) ~, 
             \quad  
   Q^{\pm\da} =\fr{1}{\sq{2}} 
         \bigl( Q^{(1)\da} \pm iM_{\da\db}Q^{(2)\db} \bigr)~,
\ee 
where $M_{\da\db}$ satisfy the orthogonal condition and 
\bb
      \gm^I_{a\da}M_{IJ}-M_{ab}M_{\da\db}\gm^J_{b\db}=0 ~. 
\ee
The solution is  given by 
$M_{ab}=(\gm^1 \cdots \gm^{p+1})_{ab}$ and 
$M_{\da\db}=(\gm^1 \cdots \gm^{p+1})_{\da\db}$
{}From the boundary condition (5.2)  $Q^{+}$ is the unbroken 
supersymmetry satisfying $Q^+ \vert B >=0$ and $Q^-$ is the 
broken one.
These satisfy the following superalgebra;
\bba
 &&  \{ Q^{+a}, Q^{-b} \}=2\twosq \momp \delta^{ab} ~, 
           \\
 &&  \{ Q^{+a}, Q^{-\db} \}=\{ Q^{-a}, Q^{+\db} \}
           =2\gm^i_{a\db}~p^i  ~, 
             \\
 &&  \{ Q^{+\da}, Q^{-\db} \}=2\twosq p^- \delta^{\da\db} ~.
\eea  
All other types of anticommutators vanish. 
The supersymmetry transformation for the unbroken supercharge 
$Q^+$ at the boundary is given by  
\bba
   &&   \hdelta^{(+)}_{\eps}X^{\a}
          = -\pconst \fr{i}{\sq{\momp}}
              \eps^{\da}\gm^{\a}_{\da a} S^{-a} ~,
            \qquad 
       \hdelta^{(+)}_{\eps}X^{\bari} =0  ~, 
           \nonumber \\
   && \quad 
           \hdelta^{(+)}_{\eps} S^{-a} 
           = \pconst \hbox{$\sq{2\momp}$} \eps^a 
            +\pconst \fr{1}{\sq{\momp}} 
               \eps^{\da} \gm^I_{\da a} \pd_C X^I  
\eea
and
\bb
     \hdelta^{(+)}_{\eps}\Bigl( \pd_C X^I \Bigr) 
       = -\spd (\hDelta^{(+)}_{\eps}X^I )~,
         \quad 
      \hDelta^{(+)}_{\eps}X^I 
           = -\pconst \fr{i}{\sq{\momp}}
              \eps^{\da}\gm^I_{\da a} S^{-a}~, 
\ee
where $\pd_C X^I =\pd X^I +M_{IJ}\bpd X^J 
=(-\spd X^{\a}~,~\tpd X^i)$.

  The vertex operators in this frame are defined at the 
boundary by
\bb
        V({\cal A}(X,\th)) =V_B(A)+V_F(\Psi) 
           = \sint {\cal A}_{\mu}(X,\th)\pd_C X^{\mu}~,  
\ee 
where $\pd_C X^{\mu}=( \tpd X^+ , \tpd X^- , \pd_C X^I )$. 
The fermionic coordinate $\th$ is now defined by 
\bb
     \th =\half \fr{1}{\pconst i\sq{\momp}}\Gm^+ S^- ~, 
\ee
where $(S^-)^t =(S^{-a},0,0,0)$ and 
$S^{-a} =\fr{1}{\sq{2}}( S^{(1)a}-iM_{ab}S^{(2)b})$.  
The superfield is given by (3.8) with the non-abelian 
generalization of  supersymmetry transformation (3.11). 
The difference is the assignment of vector field 
\bb
   A^{\mu}= \Bigl( A^{\a}(X^{\a}) ~,~ \phi^{\bari}(X^{\a}) \Bigr) ~,
\ee
where $\bari =(+,-,i)$. The fields depend only on the coordinates 
$X^{\a}$ such that $\pd_{\bari}A^{\mu}=\pd_{\bari}\Psi=0$. 
Thus $\pd^+ =0$ reduction used in the previous calculation now 
becomes essential.    

   The amplitude in the cylinder frame is defined by 
\bb
   <0\vert  V_1 \cdots V_n \vert B> 
   = g^{n-2}\sum_{perm} Tr <0 \vert 
          \int \fr{d\s_1}{i\pi}{\cal V} (\s_1) 
     \cdots \int \fr{d\s_n}{i\pi}{\cal V} (\s_n) \vert B> ~, 
\ee
where $<0\vert$ is the closed string vacuum. Although the vertex 
operators are all attached to the boundary, we need a sum over 
inequivalent permutations of the vertex operators described by 
 $\sum_{perm}$.  This just corresponds to the $\tau$-ordering 
in the open-string frame.

  As in the previous section, we can prove that the supersymmetry 
transformation for the unbroken supercharge $Q^+$ satisfies the 
equation    
\bb
      \sum_{perm} \hdelta^{(+)}_{\eps} V({\cal A})
              \exp \Bigl( -igV({\cal A}) \Bigr) 
       = \sum_{perm} V (\delta^{(+)}_{\eps} {\cal A}) 
             \exp \Bigl( -igV({\cal A}) \Bigr) ~, 
\ee
where we use the symbol $\delta^{(+)}_{\eps}$ for the 
space-time supersymmetry transformation with the field contents 
(5.12). We here show that only in the case of 
$\hdelta^{(+)}_{\eps}V_B(A)$. The variation of $V_F(\Psi)$ is 
also calculated in the same way.   
For simplicity we discuss in the special background 
$\phi^+=\phi^-=0$. In this case the variation of $V_B(A)$ is 
given by 
\bb
    \hdelta^{(+)}_{\eps}{\cal V}_B(A) 
        = {\cal V}_F (\delta^{(+)}_{\eps} \Psi)-ig Y 
             -\spd \Bigl( 
                  A^I \hDelta^{(+)}_{\eps} X^I \Bigr)~, 
\ee
where $A^I=(A^{\a}~,~\phi^i)$ and 
\bb
       Y= [A^I , A^J]\pd_C X^J \hDelta^{(+)}_{\eps} X^I 
           -\fr{i}{8}[A^K ,F^{IJ}](S^- \gm^{IJ}S^- ) 
               \hDelta^{(+)}_{\eps} X^K ~.
\ee 
Thus the variation of $V_B(A)$ in the boundary condensate  
becomes
\bba
   &&   \sum_{perm} \hdelta^{(+)}_{\eps} V_B(A) 
          \exp \Bigl( -ig V_B (A) \Bigr)  
               \nonumber \\
   && \quad  = V_F(\delta_{\eps} \Psi)-ig\sint Y  
          -ig \sum_{perm} V_F(\delta^{(+)}_{\eps}\Psi) 
              V_B(A)      
           \nonumber   \\ 
   && \qquad 
         +ig \biggl( \oint_{|z^{\pp}|>|z|}
                \fr{d z^{\pp}}{i\pi}\pd_{z^{\pp}} \Bigl( 
                A^I(\s^{\pp}) \hDelta^{(+)}_{\eps} X^I \Bigr)
                \sint {\cal V}_B(A(\s))
                        \\ 
   && \qquad\qquad  
               +  \sint {\cal V}_B(A(\s))  
                    \oint_{|z^{\pp}|<|z|} 
                    \fr{d z^{\pp}}{i\pi}\pd_{z^{\pp}} \Bigl( 
                     A^I(\s^{\pp}) \hDelta^{(+)}_{\eps} X^I \Bigr)
                 \biggr) + o(g^2) ~,
                    \nonumber 
\eea
where $z=\e^{2i\s}$ and $z^{\pp}=\e^{2i\s^{\pp}}$. 
The integral of $z^{\pp}$ gives the contribution from the 
collision point $z^{\pp}=z$. 
We here regularize the singularity by 
displacing the $\s$ contours by infinitesimal amounts along the 
cylinder axis, which is described by $|z^{\pp}|>|z|$ and $|z^{\pp}|<|z|$. 
Near the collision point $z \sim z^{\pp}$,  we take    
$z^{\pp}-z =\veps\e^{i\th}$ for $|z^{\pp}|>|z|$ and  
$z^{\pp}-z =\veps\e^{-i\th}$ for $|z^{\pp}|<|z|$,  
where $ 0 < \th <\pi$. 
The singularity of the operator product is evaluated in the 
momentum space as
\bb
     f_1(X(\s^{\pp})) f_2(X(\s)) 
       = \int dk^{\pp}dk f_1(k^{\pp})f_2(k)
            \e^{i(k^{\pp}+k)\cdot X(\s)}
         |z^{\pp}-z|^{k^{\pp}\cdot k} ~. 
\ee
So we must evaluate the following integral 
\bb
       \oint_{|z^{\pp}|>|z|}\fr{d z^{\pp}}{i\pi} 
         \pd_{z^{\pp}} |z^{\pp}-z|^{k^{\pp} \cdot k} 
       = \int_{z^{\pp}-z ~\atop =\veps \e^{i\th}} 
           \fr{d z^{\pp}}{i\pi} 
         \fr{1}{z^{\pp}-z} |z^{\pp}-z|^{k^{\pp} \cdot k} 
       =\veps^{k^{\pp}\cdot k} ~.
\ee
In the case of $|z^{\pp}|<|z|$ the integral gives the negative 
sign: $ -\veps^{k^{\pp}\cdot k}$. When $k^{\pp} \cdot k =0$ the
integral gives a finite contribution. In the coordinate space 
this just gives $ig \sint Y$, which cancels the second term of 
r.h.s. in (5.17).   

 The space-time supersymmetry $\delta^{(+)}$ is now realized linearly as  
\bb
   \delta^{(+)}_{\eps} <0 \vert V_1 \cdots V_n \vert B> 
         = <0 \vert [\eps Q^+, V_1 \cdots V_n] \vert B> =0 ~,
\ee
where the equation (5.14) is used in the first equality and 
$ <0 \vert Q^+ = Q^+ \vert B> \break =0$ in the second. 

  On the other hand the nonlinear realization of the broken 
supersymmetry $Q^-$ is given as follows. 
Let us consider a constant shift of fermionic collective coordinate  
\bb
    \delta^{(-)}_{\eta}\Psi = \eta ~. 
\ee 
If the $\eta$ satisfies the equation 
\bb
           D^{\mu} \eta =0 ~,
\ee
the vertex operator transforms under this transformation as 
\bb
      V_F(\delta^{(-)}_{\eta} \Psi) 
        = -i \sint {\bar \eta}\Gm_{\mu}\th \pd_C X^{\mu} 
        = \half \Bigl( 
            \eta^a Q^{-a} + \eta^{\da}Q^{-\da} \Bigr) ~. 
\ee
Noting the commutator $\{ Q^-,S^- \}=0$ and 
$\hdelta^{(-)}_{\eps}X^{\a}= 
\hdelta^{(-)}_{\eps}( \pd_C X^I) =0$  
such that $[Q^-,V]=0$, we obtain 
\bb
    \delta^{(-)}_{\eta} 
         <0\vert V_1 \cdots V_n \vert B> = 0 
\ee
because of $<0 \vert Q^- =0$. Thus this symmetry is exact even 
in the static gauge. The presence of the supersymmetries $\delta^{(\pm)}$ 
reflects that D-branes can couple to closed superstrings.    
Since the condition (5.22) means that $[ \phi^i , \eta] =0$ 
for multi D-branes, if we take the Chan-Paton factor is $U(n)$, 
$\eta$ is a shift in  $U(1)$ direction. 
This result seems to support the recent descriptions of  
supermembrane using multi D-particles~\cite{bfss} and 
superstring  using multi D-instantons~\cite{ikkt}.

\end{document}